\documentclass[12pt]{article}
\usepackage{epsfig}
\usepackage{eepic,overpic}

\def\beq{\begin{equation}}
\def\eeq#1{\label{#1}\end{equation}}
\def\eeqn{\end{equation}}

\def\beqa{\begin{eqnarray}}
\def\eeqa#1{\label{#1}\end{eqnarray}}
\def\eeqan{\end{eqnarray}}

\let\bar=\overbar

\def\Dslash{\not{\hbox{\kern-4pt $D$}}}
\def\dslash{\not{\hbox{\kern-2pt $\del$}}}

\def\msb{{\bar{\ssstyle M \kern -1pt S}}}

\def\Title#1{\begin{center} {\Large {\bf #1} } \end{center}}

\begin{document}

\Title{Ultra High Energy Cosmic Rays from earth-based observatories}

\bigskip\bigskip


\begin{raggedright}  

{\it Sergio Petrera \\
INFN and Physics Department, L'Aquila University\\
I-67010 L'Aquila, ITALY}
\bigskip\bigskip
\end{raggedright}

\section{Introduction}

The origin of the highest energy cosmic rays is one of the most
exciting questions of astroparticle physics. 
Even though a general concept  
linking magnetic field and size of possible
sources (the so
called ``Hillas plot''~\cite{Hillas}) is the basis of our 
current understanding, up to now there are no 
generally accepted source candidates known to
be able to produce particles of such extreme energies. 

At these energies cosmic rays are expected to exhibit a suppression
in the energy spectrum because of their interaction  with the microwave
background radiation (CMB). This feature, known as the
Greisen-Zatsepin-Kuz'min (GZK) effect~\cite{GZK},
is at about  $\sim 6 \cdot 10^{19}$ eV for protons. It limits the
horizon from which these particles can be observed to a distance 
below about 100 Mpc (depending on the
primary mass). 
The
non-observation of the GZK effect in the data of the AGASA experiment~\cite{AGASA}
has motivated several theoretical and phenomenological models trying to
explain 
the absence of
the GZK effect. 
More recently both HiRes~\cite{HiRes} and Auger~\cite{AugerSDsp} have shown
evidence of a suppression
such as expected from the GZK effect with high
statistical 
significance. 
The recent  observation of directional
correlations 
of the most energetic
Auger events with the positions of nearby Active Galactic 
Nuclei~\cite{AugerAGN} 
complements the observation of the
GZK effect very nicely.  

Mass composition is another important key to
discriminate among different 
models about the origin of high-energy cosmic rays.
Such
measurements are difficult due to their strong dependence on 
hadronic interaction
models. Only primary photons can be discriminated safely from protons 
and nuclei and recent
upper limits to their flux largely constrain existing top-down models.

In this paper, prepared for the Physics in Collisions 2008
Conference, each of these topics are exposed and reviewed.

\section{UHECR Experiments}

In the UHE region
two detection methods are effective for extensive air showers (EAS): 
arrays of surface
detectors and air fluorescence detectors. A comprehensive review
of these experimental methods can be found in~\cite{NaganoWatson}.
In this Section recent experiments dedicated to the detection of
cosmic rays are briefly described.
 
\smallskip
\noindent {\bf AGASA}. The Akeno Giant Air Shower Array, located in
Japan 
at the latitude of about 35$^\circ$ N and altitude of 900 m above sea
level 
was in operation from 1990 until 2004. It was a large surface
array~\cite{AGASANIM}, designed to measure the front of the cosmic ray
showers 
as they reach ground. The array consisted of 111 plastic scintillators 
with size of 2.2 m$^2$ deployed with separation of 1 km and covering
an area 
of 100 km$^2$. The array was complemented by 27 muon detectors
consisting 
of proportional counters placed below absorbers.

\smallskip
\noindent {\bf HiRes}. It is the new and sophisticated version of the 
pioneering Fly's Eye instrument of the Utah group based on the
detection 
of the fluorescence light from the nitrogen molecules excited by the 
charged particles of the cosmic ray showers. It was in operation from
1997
 until 2006. The HiRes instrument~\cite{HiResNIM}  consists of two
 sites 
12.6 km apart located at Dugway in Utah (USA)
hosting 22 telescopes at HiRes I and 42 at HiRes II. The telescopes
cover
 the full 360$^\circ$ azimuth and in elevation from 3$^\circ$ up to
 17$^\circ$ 
(HiRes I) and from 3$^\circ$ up to 31$^\circ$ (HiRes II). 
The main components of each telescope are a spherical mirror of about
4 m$^2$ 
size and an array of 256 photomultipliers as sensitive element.
UV filters to cut light outside the 300-400 nm interval of the
nitrogen 
fluorescence were also used.

\smallskip
\noindent {\bf Auger}. Two observatories, one in the Northern and one
in the Southern hemisphere are foreseen for the Pierre Auger
Observatory project, 
to achieve a full exploration of the sky.
The Southern Auger Observatory~\cite{AugerNIM} is located near the
small town 
of Malarg\"ue in the province of Mendoza (Argentina) at the latitude
of about 
35$^\circ$ S and altitude of 1400 above see level. 
The Observatory is a hybrid system, a combination of a large surface
array 
and a fluorescence detector.
The surface detector (SD) is a large array of 1600 water Cherenkov
counters 
spaced at a distance of 1.5 km and covering a total area of 3000
km$^2$. 
Each counter is a plastic tank of cylindrical shape with size 10 m$^2 \times
1.2$ m 
filled with purified water. The SD tanks activated by the event record
the 
particle number and the time of arrival. From the times, the direction 
of each event is determined with an accuracy of about 1$^\circ$. 
The fluorescence detector
(FD) 
consists of 24 telescopes located in four stations which are built on 
the top of small elevations on the perimeter of the site. The
telescopes 
measure the shower development in the air by observing the
fluorescence light. 
Each telescope has a 12 m$^2$ spherical mirror with curvature radius
of 3.4 m 
and a camera with 440 photomultipliers. The field of view of each 
telescope is 30$^\circ \times 30^\circ$. UV filters were used as in HiRes.
The Southern Auger Observatory started to collect data in 2004. The
Observatory and has been completed in Summer 2008. 
The Northern Auger Observatory which is now being designed will be 
located in Colorado (USA).
We note that the present Auger Observatory is the only detector
exploring the Southern hemisphere. 

\smallskip
\noindent {\bf Telescope Array (TA)}. It is being built by a
US - Japan - Korea 
collaboration in Millard County, Utah, USA.  Like the Auger Observatory, 
the TA 
is a hybrid detector~\cite{TA}. 
It covers an area of
860 km$^2$ and comprises 576 scintillator stations and three FD sites
on a 
triangle with about 35
km separation each equipped with 12 fluorescence telescopes.
\smallskip

\begin{figure}[!ht]
\begin{center}
\epsfig{file=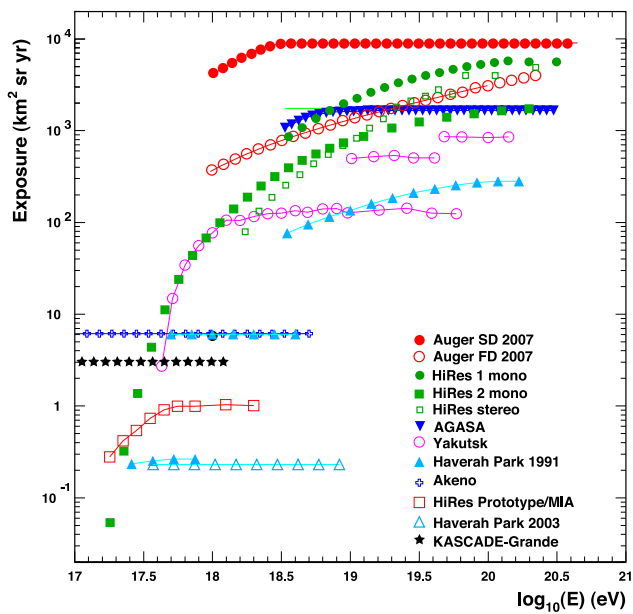,height=8cm}
\caption{\small{Accumulated exposures of various experiments at the end of 2007 
(see ref.~\cite{Kampert, Bergman}).}}
\label{fig.expo}
\end{center}
\end{figure}

Fig. \ref{fig.expo} shows a comparison of the exposures accumulated
by various experiments at the end of 2007. More details about the exposure
calculations
can be found in~\cite{Kampert, Bergman}. It can be seen that the largest exposure 
has been achieved with the Auger Observatory
and it will continue to deliver more than about 7000 km$^2$ sr for
each year of operation. TA is not shown in the figure since it started
full operation since March 2008.  
It is important to notice the different behaviour with energy of
the apertures for arrays of surface detectors and fluorescence detectors. 
In case of arrays of detectors with a regular pattern, the aperture can be
calculated in a straight forward and model
independent way, once the energy threshold for CR detection and
reconstruction 
is exceeded.
The situation is different for fluorescence telescopes. Here, the maximum
distance out to which showers can be observed increases with
increasing 
fluorescence light and thereby increasing energy.  This condition
makes the aperture calculation dependent on Monte Carlo simulation
and then on primary mass and on the hadronic interaction models
employed.
This dependence can be considerably reduced by applying quality cuts
to geometry parameters (e.g. the distance of the shower), but this is
possible only if geometry is well determined as in the cases of hybrid or
stereo detection.

\section{The Energy Spectrum}

\begin{figure}[!htb]
\begin{center}
\epsfig{file=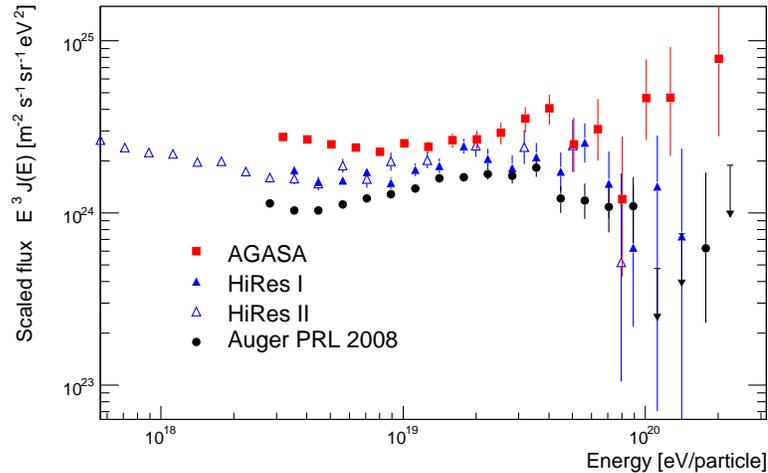,width=.7\textwidth}
\caption{\small{ Cosmic ray flux measurements
    (multiplied by E$^3$) from AGASA~\cite{AGASA}, HiRes~\cite{HiRes}
  and Auger~\cite{AugerSDsp}.}}
\label{fig.flux}
\end{center}
\end{figure}

Most of the  energy spectrum data available today\footnote{
Spectrum data from instruments with exposure less than 1000 km$^2$ sr yr
have not been considered in this review.
}
 at UHE are provided by 
AGASA, HiRes and Auger (see Fig. \ref{fig.flux}). The two last
experiments 
recently published
spectrum analyses~\cite{HiRes,AugerSDsp} showing evidence of a 
flux suppression as expected by the GZK effect with significances 
of about 5 and 6 $\sigma$ respectively at slightly different energies 
(5.6 and 4 $\times 10^{19}$ eV). Shifting the energy scale by about
+15\% for Auger and about -25\% for AGASA with respect to HiRes the
three spectra agree rather well up to about 5 $\times 10^{19}$ eV. At
higher energies the AGASA data do  not exhibit any flux suppression
and thus are inconsistent with the other data.

Typical uncertainties of the energy scale are on the order of 20 $\div$
25\%. Ground arrays like AGASA rely entirely on EAS simulations with
their
 uncertainties originating from the limited knowledge of hadronic 
interactions. Fluorescence telescopes, such as operated by HiRes and
Auger, observe the longitudinal shower development in the atmosphere. 
In this way, the atmosphere is employed as a homogenous calorimeter. 
Nonetheless possible differences in their energy recostruction are
still present because of different assumptions (e.g. fluorescence
yield, event reconstruction, analysis cuts).

Even though the Auger spectrum in~\cite{AugerSDsp} is based on
surface data, the energy calibration is quite new. In fact 
the method used by Auger to measure the energy spectrum exploits the 
hybrid nature of the experiment with the aim of using the data itself 
rather than simulations.
For each event the tanks of the SD measure the particle density
expressed in units of VEM (Vertical Equivalent Muons) and the times of 
arrival which are used to determine the axis of the shower. The
dependence 
of the particle density on the distance from the shower axis 
is fitted by a lateral distribution function (LDF). The LDF fit allows 
determining the particle density S(1000), expressed in units of VEM, at 
the distance of 1000 m from the axis. This quantity is a good energy 
estimator~\cite{Newton} in the sense that it is strongly correlated with the 
energy of the cosmic ray and almost independent of the mass.
The energy estimator S(1000) depends on the zenith angle because of
the atmosphere attenuation. The value of S(1000) corresponding to 
the median zenith angle of 38$^\circ$ (S$_{38}$) is used as reference 
and the zenith 
angle dependence of the energy estimator is determined assuming that 
the arrival directions are isotropically distributed. This procedure 
is traditionally called {\it Constant Intensity Cut}.
\begin{figure}[!htb]
\centering
\begin{tabular}{ll}
  \begin{overpic}[width=0.38\textwidth]{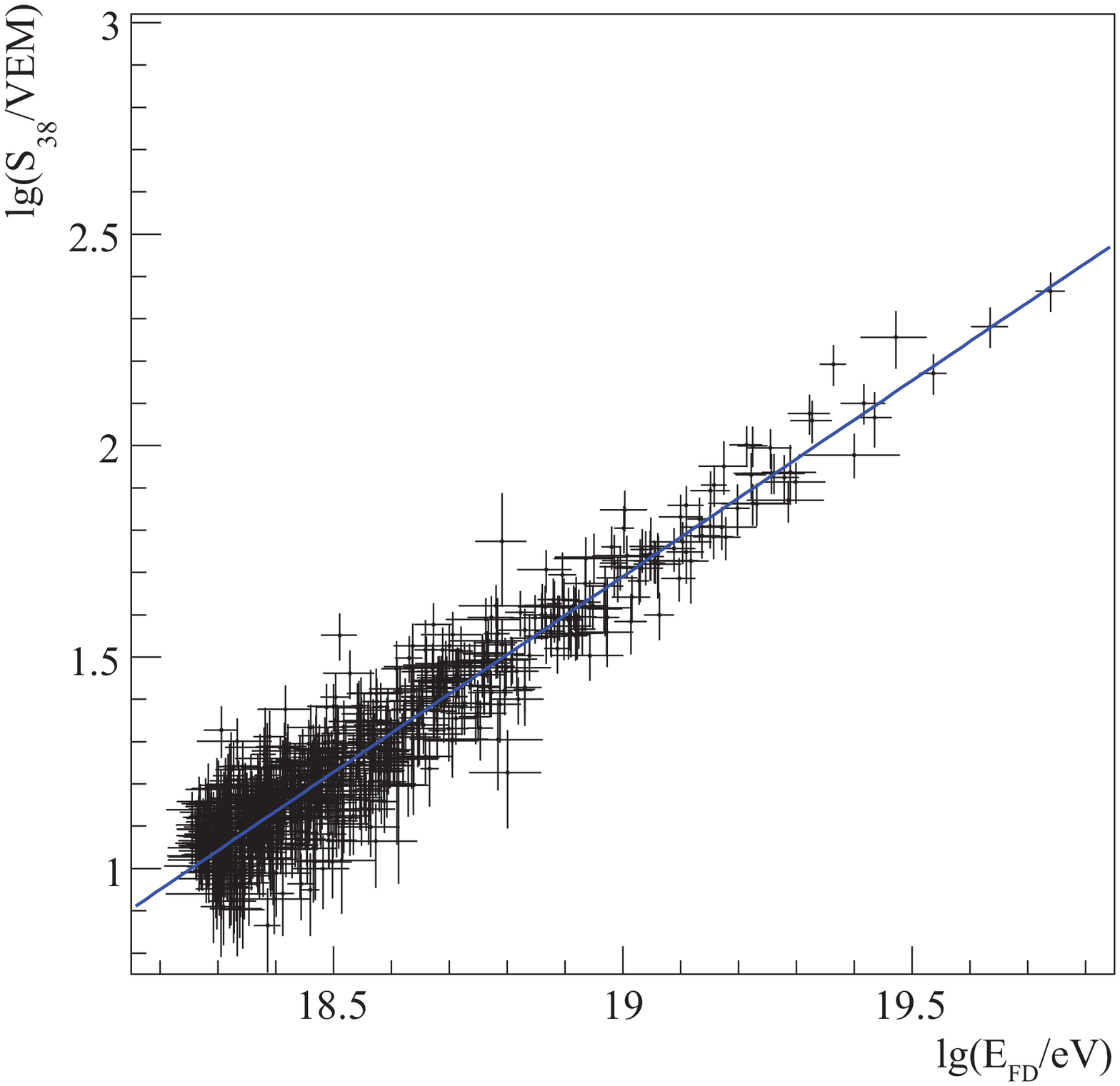}
    \put(13.8,50.5){\includegraphics[width=0.18\textwidth,clip=]{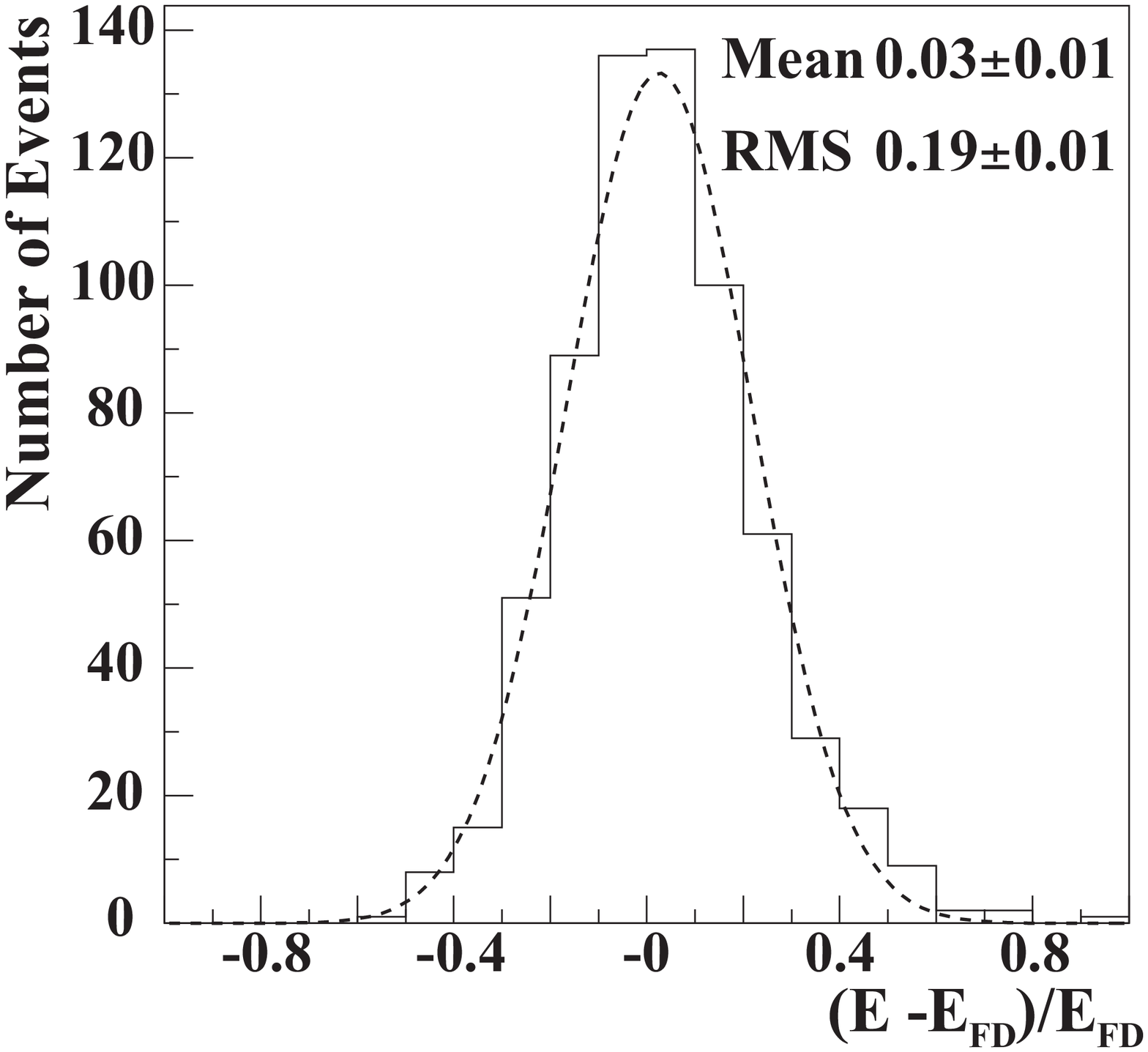}}
  \end{overpic}
&
\includegraphics*[width=.62\textwidth]{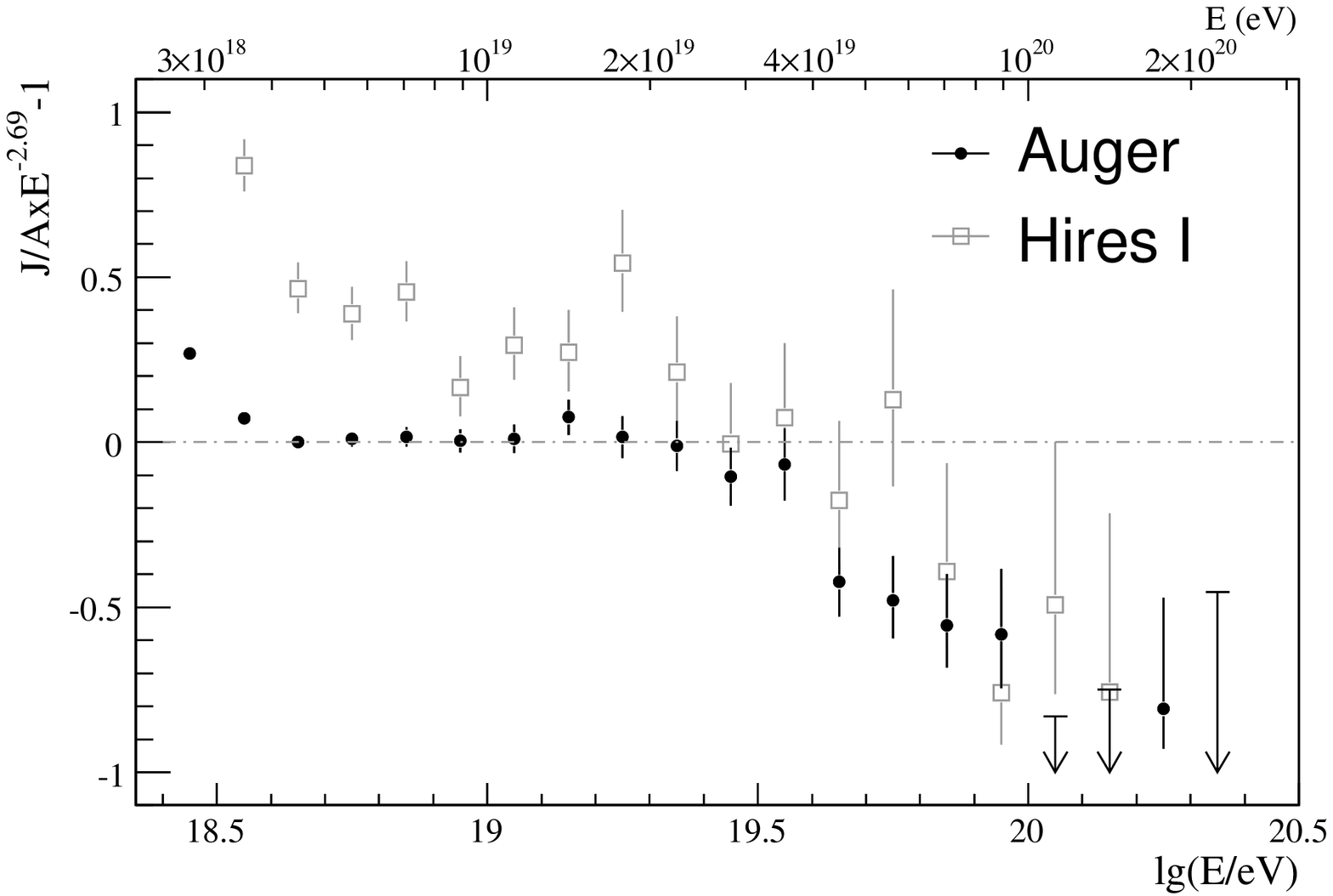}
\end{tabular}
\caption {\label{fig.AugerPRL}{\small{\it Left}: Auger calibration of SD
  data: correlation between surface
  detector signal and FD energy. The
 fractional differences between the two energy estimators are inset. 
 {\it Right}: Fractional difference between Auger and HiRes I data
relaltive to a spectrum with index of 2.69.}}
\end{figure}
The absolute calibration of S$_{38}$ is derived from the hybrid events 
using the calorimetric energy measured by the FD which is then
corrected 
for the missing energy using the mean value between proton and iron 
(uncertainty about 4\% at 10$^{19}$ eV). This absolute calibration, which 
defines the energy scale, is at present affected by a systematic error
 of about 20\%, mainly due to uncertainties on the fluorescence
 yield 
and on the calibration of the FD telescopes.
The energy calibration, obtained from the subset of hybrid events 
(see Fig. \ref{fig.AugerPRL}) is then used for the full set of 
events with higher
 statistics measured by the SD.

The flux suppression in Auger and HiRes as well as the possible
difference in their energy scales is evident when plotting the
fractional difference with respect to a power law spectrum.
Fig. \ref{fig.AugerPRL}, right panel, shows this fact for a spectral
index of 2.69 which is the one fitted by Auger below  4 $\times 10^{19}$ eV.

\section{Primary Composition}

\begin{figure}[!htb]
\begin{center}
\epsfig{file=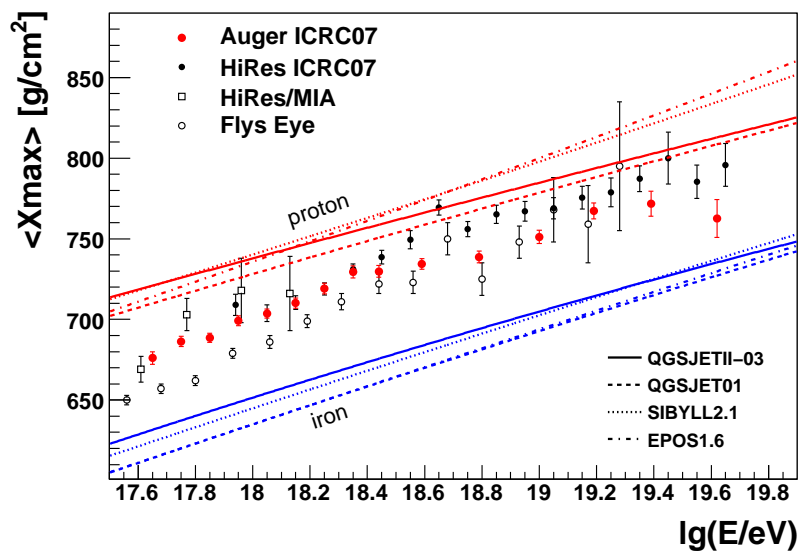,width=.75\textwidth}
\caption{\small{$\langle X_{max} \rangle$ as a function of energy compared with proton
and iron predictions using different hadronic interaction models.}}
\label{fig.ER}
\end{center}
\end{figure}
Measuring the composition of cosmic rays is crucial
to obtain a full understanding of their acceleration
processes, propagation and relation with
galactic particles. The atmospheric depth $X_{max}$
denotes the longitudinal position of the shower
maximum, which is directly accessible with the FD.
It grows logarithmically with the energy of the primary
particle. The behaviour of $X_{max}$ for different
primary particles like photons, protons and heavier
nuclei can be conceptually understood in the
framework of the Heitler and superposition models~\cite{Matthews}, 
which provides  good
agreement with detailed Monte Carlo simulations.
New results based on HiRes-Stereo and Auger hybrid data 
 at the ICRC~\cite{HRICRC,AugICRC} are reported in 
Fig. \ref{fig.ER}.
Both data sets agree very well up to $\sim 3 \cdot 10^{18}$ eV
but differ slightly at higher energies. The differences between 
the two experiments are within the differences observed 
between p- and Fe-predictions for different hadronic
interaction models. With these caveat kept in mind, both experiments 
observe an increasingly lighter composition towards the ankle. 
At higher energies, the HiRes measurement yields a
lighter composition than Auger. 

\begin{figure}[!htb]
\centering
\begin{tabular}{ll}
\includegraphics*[width=.48\textwidth]{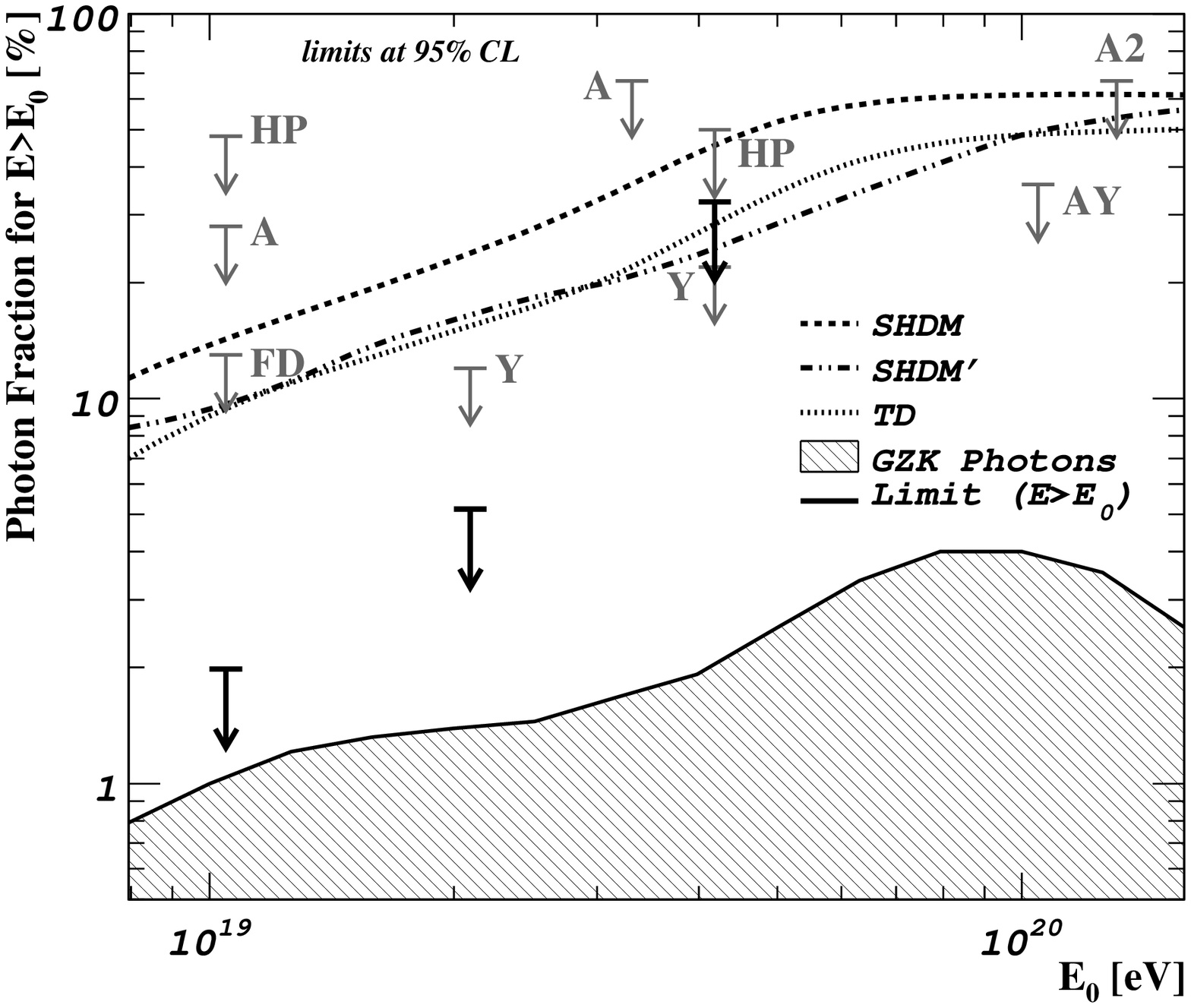}
&
\includegraphics*[width=.45\textwidth,height=.38\textwidth]{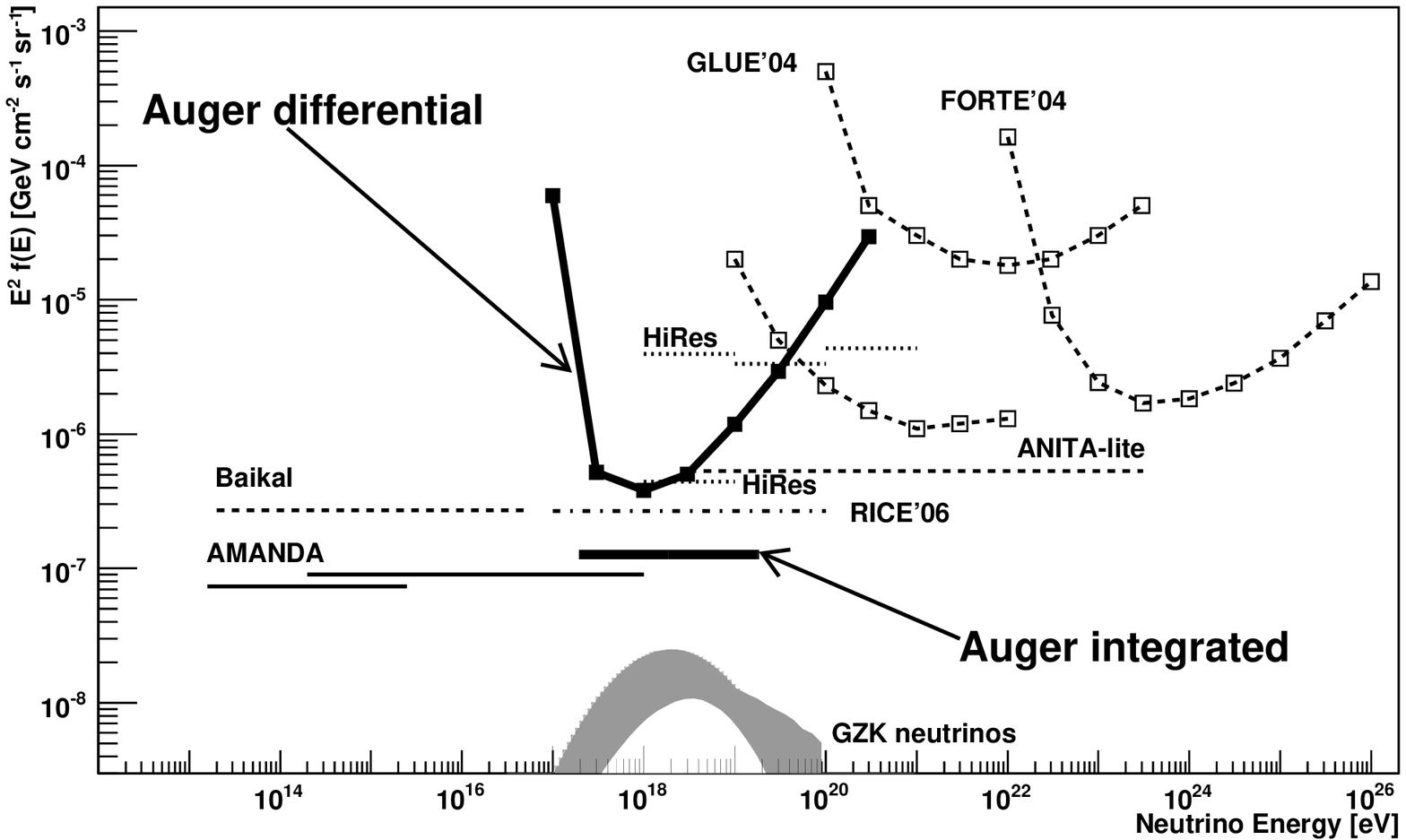}
\end{tabular}
\caption {\label{fig.photneut}{{\small \it Left}: The upper limits on the 
fraction of photons in the integral cosmic ray
flux derived from Auger SD (black arrows) along with previous
experimental limits (HP: Haverah Park; A1, A2: AGASA;
AY: AGASA-Yakutsk; Y: Yakutsk; FD: Auger hybrid limit).
Also shown are predictions from top-down models (SHDM, SHDM', TD) 
and predictions of the GZK photon fraction. For references 
see ~\cite{photSD}. 
 {\it Right}: Limits at the 90\% C.L. for a diffuse flux of
 $\nu_{\tau}$ assuming a 1:1:1 ratio of the 3 neutrino flavors and
 the expected flux of GZK neutrinos.
For references see~\cite{neutrinos}}}
\end{figure}

Another important issue concerning the primary composition is the search
for photons and neutrinos in EAS. The Auger Observatory has set 
new photon limits with
both the hybrid and SD detection methods~\cite{photHy,photSD}.
The new limits are compared to previous results and to
theoretical predictions  in
Fig. \ref{fig.photneut} for the photon fraction.  In terms of the photon
fraction, the current bound at 10 EeV approaches the percent
level while previous bounds were at the 10 percent level.
A discovery of a substantial photon flux could have been
interpreted as a signature of top-down (TD) models. In turn, the
experimental limits now put strong constraints on these
models. For instance, certain SHDM (Super Heavy Dark Matter) 
or TD models discussed
in the literature~\cite{phmodels}  predict fluxes that exceed the limits
by a factor 10. 

 Neutrino induced showers can be also
identified if they occur deep in the atmosphere under
large zenith angles, or by their special topology
in the case of Earth-skimming tau neutrinos.
Identification criteria have been developed to find
EAS that are generated by tau neutrinos emerging
from the Earth. Auger has searched for tau neutrinos in
the data collected up to
August 2007. No candidates have been found and an upper limit 
on the diffuse tau neutrino flux has been set. In Fig.
\ref{fig.photneut} 
this result~\cite{neutrinos} is shown.

\section{Arrival Directions}

Most of the recent results are from the Auger Collaboration who have
started a detailed investigation of the angular directions of the
cosmic rays. While no excess has been found from the Galactic Centre
in the EeV energy range, evidence for anisotropy has been found in the
extreme energy region.

Observation of an excess from the region of the Galactic Centre in
the EeV energy region  were reported by AGASA~\cite{AGanis} and 
SUGAR~\cite{SUanis}.
The Auger Observatory is suitable for this study because the Galactic
Centre (constellation of Sagittarius) lies well in the field of view 
of the experiment. The angular resolution of the SD of Auger depends 
on the number of tanks activated by the shower and it is better than 
one degree at high-energy. However, with statistics much greater than
previous data, the
Auger search~\cite{Auanis} does not show abnormally over-dense regions 
around the GC.

\begin{figure}[!htb]
\begin{center}
\epsfig{file=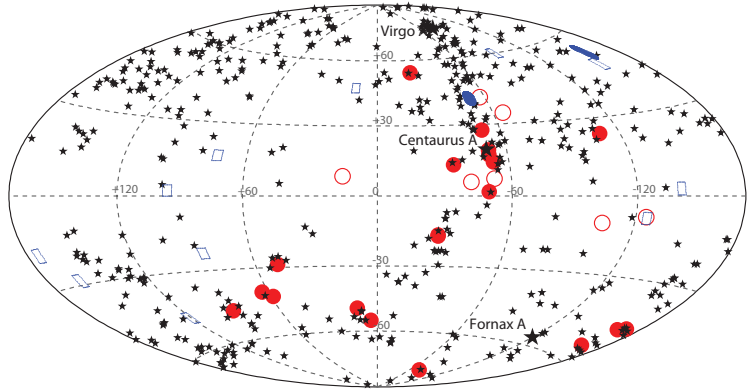,width=.75\textwidth}
\caption{\small{The sky seen with UHECRs with energy above 56 EeV detected
  with the surface array of Auger (red
  circles,~\cite{AugerAGN,AugerAGNAp}) and with the HiRes detector in
  stereo mode (blue squares,~\cite{HiResAGN}) in galactic coordinates.
 Filled markers denote cosmic rays within 3.1$^\circ$ from AGNs with
 redishift $z<0.018$ (black stars,~\cite{VCV}). The relative exposures
 of the two experiments are not shown for simplicity. Very roughly
 Auger (HiRes) is blind to a part of the left (right) side of
this plot and then their exposures are rather complementary.  
Detailed exposures can be found in the original papers.}}
\label{fig.skymap}
\end{center}
\end{figure}

A big step towards the discovery of the UHECR sources has been
recently made by the Pierre Auger Collaboration~\cite{AugerAGN,
  AugerAGNAp}. The highest energy events recorded so far were scanned 
for correlations
with relatively nearby AGNs ($z \leq 0.024$ corresponding to $D
\leq 100$ Mpc) listed in the
V\'eron-Cetty/V\'eron catalogue~\cite{VCV}. AGNs where used
only up to a maximal redshift $z_{max}$, which was a free parameter
in the correlation scan. Two other free parameters were the minimal
energy of the cosmic ray events E$_{thr}$ and the maximum
separation between reconstructed cosmic ray direction and the AGN
position $\psi$. The scan was performed over data taken
during the first two years of stable operation (01/2004 - 05/2006) and
a significant minimum of the chance probability calculated assuming
isotropic arrival directions was observed. After the parameters of this
{\it explorative scan} ($z_{max}=0.018$,
E$_{thr}= 56~EeV$, $\psi=3.1^\circ$) were fixed, the
consecutive data set (06/2006-08/2007) was used to verify the
correlation signal and the hypothesis of an {\it isotropic source
  distribution} could be rejected at more then 99\%
confidence level. A sky map of the $27$ events above the energy
threshold of E$_{thr}= 56~EeV$ together with the selected
AGN is shown in Fig.~\ref{fig.skymap}. Also shown are
the events selected during a follow-up analysis of stereo data from
the HiRes Collaboration~\cite{HiResAGN}, which do not
show a significant correlation. 

The interpretation of the observed anisotropy is ongoing and a much
larger event statistics will be needed to investigate, for
example, whether the AGNs act only as tracers for the underlying true
sources and whether the angular separation between AGN and UHECR can be
related to magnetic deflections.

\section{Conclusion and outlook}

In recent years UHECRs have shown a variety of exciting features: the
flux suppression at energies as the one expected for the GZK cutoff
and possible correlations with sources are the most attactive. The two 
phenomena are strictly related one to each other. In particular the
correlation scenario is compatible with suitable
spectrum shapes and mass compositions in the GZK region. This because 
cosmic ray propagation through galactic fields and their
interactions with the photon background affect not only directions,
but also the energy and type of particles observed on Earth. 

Coming years are expected to be fruitful. New data will come from the 
Northern Hemisphere: Telescope
Array, now, and Auger North, in a few years, will join this
fascinating exploration.



\begin{thebibliography}{99}
\bibitem{phmodels} References to theoretical models can be found
  e.g. in~\cite{photSD}.
\end{thebibliography}
\end{document}